# Quelle place pour les arcs électriques et les réacteurs plasmas dans l'« Inventaire et valorisation des collections, archives scientifiques et biens culturels » de l'université de Limoges ?

Anne-Marie DELAUNE[1*], Bernard PATEYRON[2]

[1] Mission Patrimoine scientifique et culturel Université de Limoges, 7 rue Félix Eboué 87000 Limoges
[2] Sciences des matériaux céramiques et traitements de surface CNRS UMR6638, 12 rue ATLANTIS CEC 87078 Limoges Cedex
[*](auteur correspondant : anne-marie.delaune@unilim.fr)

**Résumé –** L'Université de Limoges dans le cadre de la réhabilitation de ses locaux scientifiques s'est trouvé confrontée à la mise au rebut de biens patrimoniaux irremplaçables dans le contexte « post moderne » ou « post industriel » de notre société industrielle. C'est pourquoi une « Mission Patrimoine scientifique et culturel » a été créée en 2010 afin de sauvegarder le plus possible la mémoire du dernier demi-siècle. Pour des raisons que nous évoquons les plasmas et les arcs électriques ont eu une place importante à cette époque dans l'industrie occidentale et fortuitement particulièrement à Limoges

## 1. Introduction

C'est une banalité de constater que quel que soit le jugement de valeur que nous portions sur elles, ces cinquante dernières années une période particulièrement exceptionnelle de l'histoire de l'humanité : trente glorieuses ou gaspillage des ressources. Il est également plus que probable que de telles conditions ne se retrouveront plus avant longtemps. Puisque sociologues et philosophes prétendent que nous quittons à travers une addition de crises séparées (crise économique, crise financière, crise écologique, crise scientifique, etc.) d'une civilisation industrielle à une civilisation postmoderne[1,2].

Faut-il conserver la mémoire de ce moment ? Certainement oui, puisque *« la condition de tout progrès humain est fortement liée au dynamisme de la mémoire collective »* et afin de nous éviter *« de continuels retours... en arrière et de redécouvrir la roue tous les 25 ou 50 ans »*.

C'est pourquoi, localement, la « Mission Patrimoine scientifique et culturel de l'Université de Limoges » a été créée en 2010 pour faire l'« Inventaire et valorisation des collections, archives scientifiques et biens culturels » de l'université.

### 1.1. La Mission Patrimoine scientifique et culturel de l'Université de Limoges

Elle a pour objet de constituer la mémoire de l'université (à l'exclusion des archives de l'administration). Elle se propose donc de :
- Collecter et sensibiliser au patrimoine matériel et immatériel,
- Sauvegarder et rechercher les moyens de sauvegarde
- Valoriser et faire alliance avec des structures de valorisation.

Le patrimoine proprement dit est matériel ou immatériel.
- Matériel avec les instruments et appareils, les écrits, les documents des chercheurs, tiré-à-part, cahier de labo, images, sons, supports multimédia, etc.
- Immatériel avec les pratiques, représentations, expressions, connaissances et savoir-faire, mémoire humaine (« souvenirs » à interroger), etc.

Dans ce cadre général le laboratoire Réacteur plasma (120 m2 et 860 kVA), implanté en 1977 sous l'égide du CEA (CEN Saclay), EDF et Pechiney Électrométallurgie constitue un élément exceptionnel en ce qu'il est la dernière installation opérationnelle de ce type à témoigner de la politique énergétique et industrielle de ces cinquante dernières années. Ce laboratoire est le dernier héritier :

- des travaux effectués de 1960 à 1970 dans le cadre de la filière Magnétohydrodynamique (MHD)[3].
- Des efforts et investissement dans la filière électrométallurgique et métallurgie plasma à la suite des mutations :
  - concentration de l'industrie métallurgique dans les ports : Fos, Dunkerque et abandon de la sidérurgie lorraine.
  - mise en place du programme de production électrique « tout nucléaire » d'Electricité de France

## 2. Le contexte technologique général : les grandes évolutions mondiales

### 2.1. Une énergie peu chère mais à coût fluctuant

Au cours des cinquante dernières années l'énergie est restée à faible coût bien que ces couts aient subis des variations brutales, qui ont été appelées « choc pétrolier ». Par contre la qualité de cette énergie a beaucoup varié avec la disparition des produits solides (houilles, coke, etc.) au profit de combustibles légers liquides, gazeux et même premier projets hydrogène (ainsi la RCP Hydrogène du CNRS de M. GLEIZER).

### 2.2. La pression écologique

La distribution des gaz fossiles a induit la disparition des usines de production de gaz de ville. Ainsi le sous-produit coke, combustible des fours métallurgique, est devenu rare, puisque la pression écologique interdisait de fait la construction de nouvelles cokeries polluantes et sales.

### 2.3. Essai d'un nouveau modèle économique de la métallurgie mondiale

Tentatives d'études d'un modèle économique alternatif à la concentration et au gigantisme, celui d'une électrométallurgie distribuée en micro installations de production sur les sites consommateurs. Solutions qui étaient en outre moins couteuses en investissements.

### 2.1. Les grandes restructurations nationales de 1977

#### 2.1.1. Abandon de la filière Magnétohydrodynamique (MHD)

La filière MHD avait été développée afin de pallier au fait que les turbines à gaz ne pouvaient accepter des gaz de température supérieures à 1400°C pour des raisons de tenue des matériaux et que ce fait le rendement de Carnot était inutilement dégradé. Il s'agissait de rendre conducteurs à hautes températures les gaz de combustion en les ensemençant avec des carbonates de métaux alcalins et de le faire débiter sous un fort champ magnétique dans un réacteur du type de la Figure 1. Il restait à réaliser des électrodes susceptibles de collecter le courant induit dans des tuyères à température de paroi élevée (supérieures à 1600°C) [4].
C'est pourquoi le réacteur tournant (Figure 2) avait été développé. Il était issu d'une collaboration entre une équipe du Centre d'Études Nucléaires de Saclay, de la société belge ARCOS qui était leader mondial sur le marché des électrodes enrobées de soudage à l'arc électrique et du Laboratoire CNRS du four solaire d'Odeillo.

La chute des cours du pétrole fit abandonner la filière MHD et engagea EDF et le CEA dans une réflexion sur la politique énergétique de la France. C'est à cette occasion que furent créés des groupes d'idées et de veille technologique : les « Clubs EDF » dont le Club Hautes Températures en décembre 1975. **C'est de cette réflexion que naquit l'implantation d'une activité plasmas de puissance à Limoges dans le contexte général que nous décrivons ici**

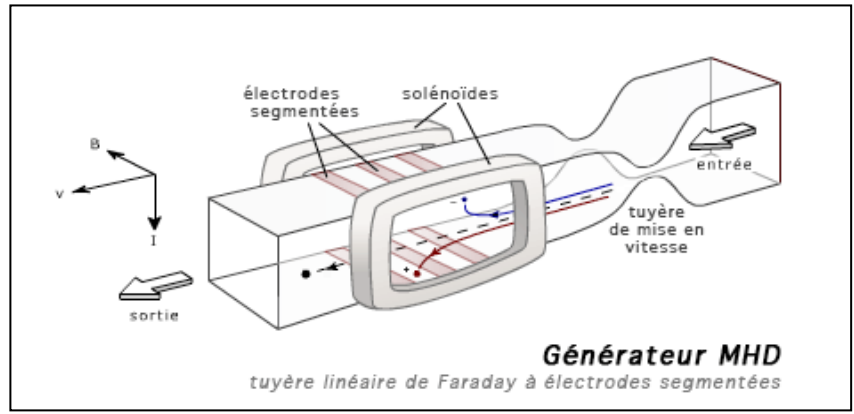

*Figure 1- Schéma de principe d'un générateur Magnétohydrodynamique de type Faraday [3].*

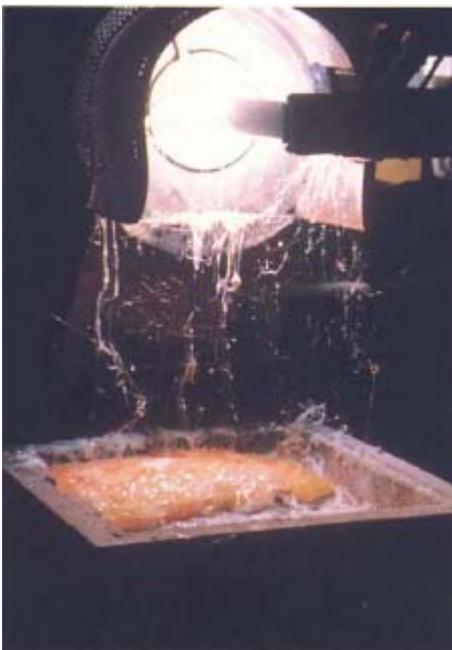
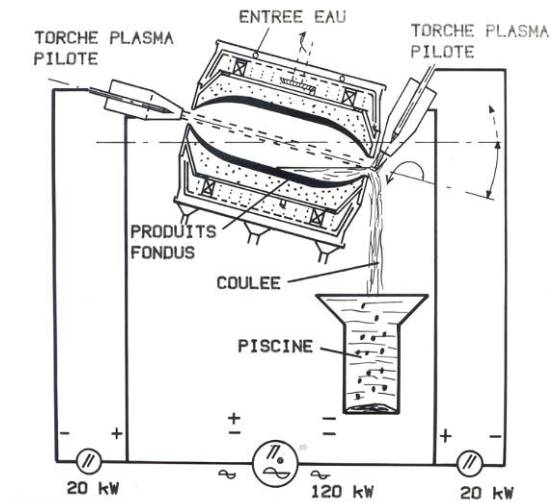

*Figure 2- Le réacteur plasma rotatif développé en collaboration par le CNRS (Foex, Delmas à Odeillo) [5] et le CEA (Yerouchalmi) [6], photographie d'une coulée d'alumine électro fondue. A droite schéma de principe du réacteur. Les fours rotatifs à arc transféré furent alors utilisés à la synthèse de céramiques électro fondues et à la production de poudres ultrafines. Certaines de ces poudres étaient alors candidates à la constitution de barrières de diffusion gazeuse pour séparer les isotopes d'uranium transportés sous forme hexafluorure. En 2007, un tel réacteur est en service au centre CEA de Cadarache. Il était utilisé tant pour l'étude du confinement vitreux de rejets de l'industrie nucléaire que pour simuler la fusion de cœur de centrale nucléaire.*

### *2.1.2. L'énergie à faible coût, le passage à l'électricité «tout nucléaire» et la restructuration de la sidérurgie française*

En 1975, le pétrole est abondant et les dernières exploitations nationales de houilles sont abandonnées. Electricité de France qui pressent les « chocs pétroliers » recherche l'indépendance énergétique et met en place le « tout nucléaire », c'est-à-dire un socle de production électrique assurée à plus de 80% par des centrales nucléaires. (Cf. le Discours du contrôleur général Fabre de décembre 1976 [7]) et elle met en place des groupes de réflexion ou club déjà cités.

Dans le même temps la sidérurgie lorraine qui exploite des minerais de fer, impurs à faibles teneurs et producteurs de scories phosphoriques, est abandonnée pour une sidérurgie portuaire. Celle-ci à Fos et à Dunkerque utilise les minerais de fer de Scandinavie ou de Mauritanie.

L'électrométallurgie était le créneau logique d'utilisation de l'énergie nucléaire d'étiage et d'heures creuses. Au sein de l'électrométallurgie les procédés plasmas ouvraient la perspective de procédés innovants et la levée de verrous technologiques.

L'implantation d'une plateforme plasma pilote, sur la base de la délocalisation du laboratoire du CEN Saclay, fut décidée et après des hésitations entre le site de Renault et celui de l'université de Lille c'est finalement Limoges qui fut choisi. Ce qui permit à l'université de Limoges d'accueillir le $3^{eme}$ Congrès IUPAC de Chimie des plasmas[1] en juillet 1977. C'était la reconnaissance de fait du nouveau laboratoire par la communauté internationale. Les actes de ce congrès rendent compte de l'état de l'art des arcs électriques en 1977.

## 3. Les activités Plasmas et arcs électriques de puissance à Limoges.

Le bâtiment Réacteur Plasma fut livré en avril 1977 ce qui permit alors les premières réalisations plasma de puissance.

### 3.1. La métallurgie extractive:

#### Métallurgie des ferroalliages (1977 À 1981)

Les activités furent centrées sur l'étude de la métallurgie plasma du ferrochrome en collaboration avec SOFREM (devenue Péchiney Électrométallurgie). Cette recherche se poursuivit jusqu'à l'abandon de la production des ferrochromes par SOFREM, en raison des difficultés d'approvisionnement liées à une politique de prix dissuasive pratiquée par les deux principaux fournisseurs, l'un et l'autre sous embargo: l'Afrique du Sud en raison de sa politique d'apartheid et l'URSS en raison de la guerre froide.
Les réacteurs conçus dès 1981, en particulier, avec l'innovation de la cathode plongeante, sont à l'origine de nombreux réacteurs plasma dans le monde parmi ceux-là, le réacteur "Plasmacan" de la société Noranda (Canada) exploité par la Société David McKee en Australie (Figure 5) et en Afrique du Sud pour le Mintek [8]. Parallèlement dès 1979 les premiers essais de traitement de MoS2, débutèrent sur le réacteur plasma à Limoges.

---

[1] Tous les deux ans, ce congrès se tient usuellement dans de grandes métropoles, en 2011, c'est le *20th International Symposium on Plasma Chemistry* du 24 au 29 juillet 2011 à Philadelphie au USA
ISPC est une conférence internationale biannuelle qui présente l'état des avancées les plus récentes dans tous les domaines de la chimie des plasmas et de ses applications.

# Métallurgie de l'aluminium - Réactualisation du procédé ALCAR (1982 à 1984)

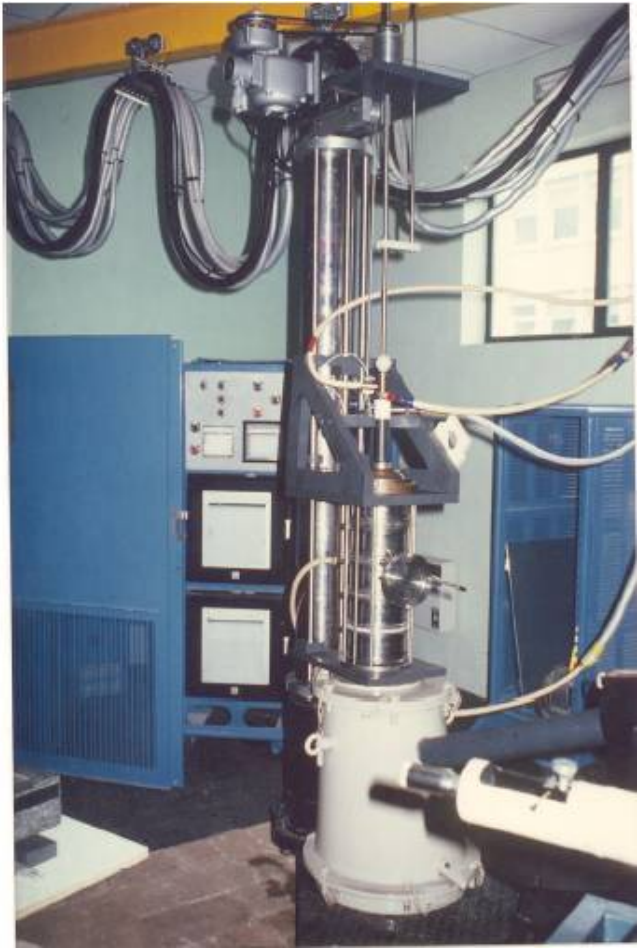

Figure 3 - *Le premier réacteur conçu sur une idée de Betheleem Steel (USA) [9]: le film ruisselant, modifiée avec l'innovation de la cathode plongeante (1978).*

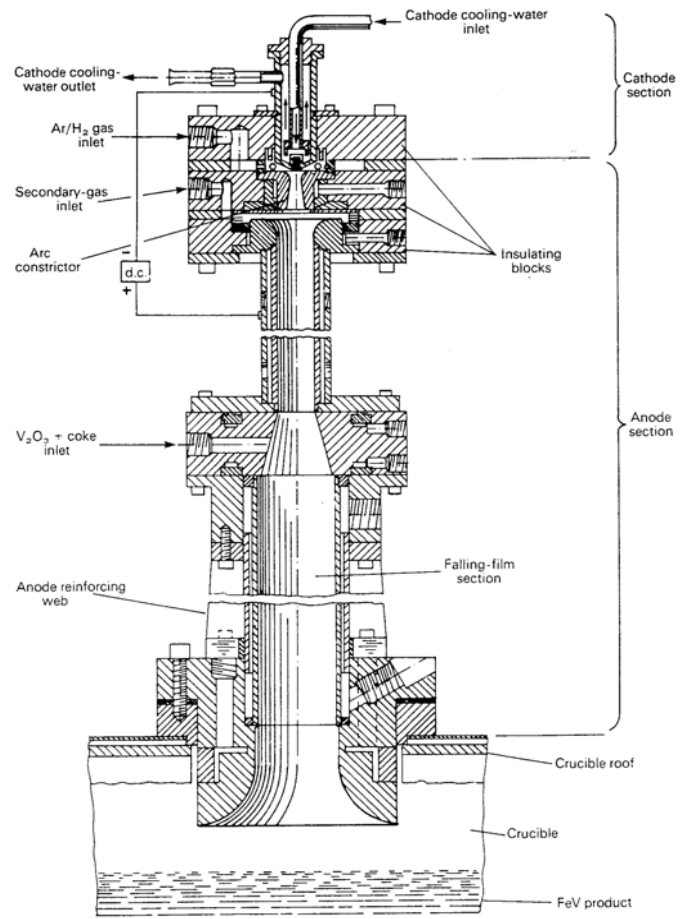

Figure 4 - *Le réacteur à film ruisselant de Betheleem Steel 1MW (1977) selon MacRae [10]*

En collaboration contractuelle, confidentielle, avec Aluminium Péchiney était entreprise l'étude des réactions de carboréduction de l'alumine (procédé ALCAR). Ce procédé de réduction directe de l'alumine en four à arc, par la voie carbure, avait été abandonné vers 1950, alors qu'il était exploité dans une usine pilote. La réaction devenait possible avec le plasma qui évitait le gigantisme des unités d'électrolyse. Il s'agissait de vérifier si la mutation de ce procédé avec le remplacement du four à arc à électrode de graphite par un plasma d'arc transféré, rendait son exploitation économiquement viable.

### 3.2. Métallurgie d'affinage (1985 À 1988)

En collaboration avec la Société du Ferromanganèse Paris-Outreau (SFPO), étude d'un procédé de métallurgie d'affinage, la décarburation et le désiliciage du ferromanganèse. Cette société produisait du ferromanganèse en hauts fourneaux et ce produit était carburé et silicié. Il existait un marché pour un produit à forte valeur ajoutée, exempt de carbone et de silicium. La réaction entre le métal carburé et son oxyde (minerai) permet de le décarburer en libérant CO et d'éliminer le silicium en relâchant $SiO_2$ dans le laitier. Un four pilote à plasma d'arc de

500 kW fut installé à Outreau près de Boulogne-sur-Mer. L'ingénieur responsable, Mlle Françoise LERROL, et le conducteur du réacteur furent formés trois ans à Limoges.

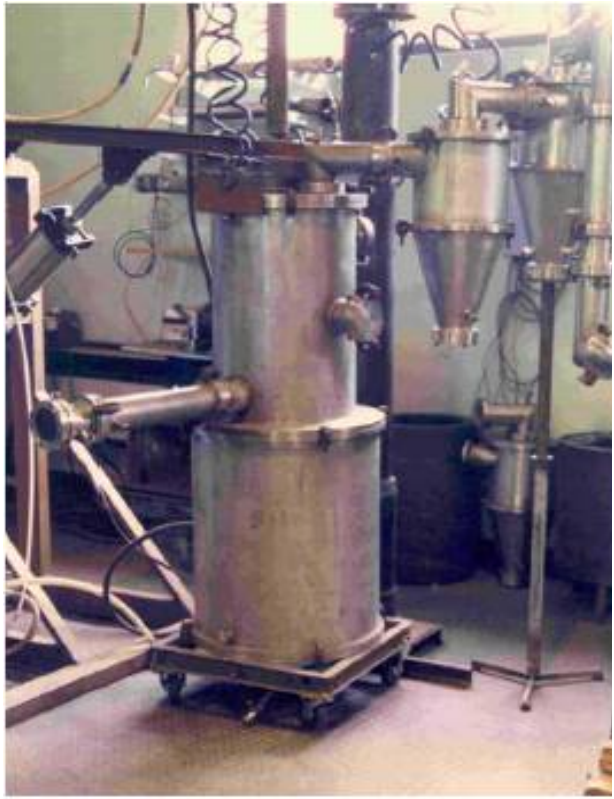
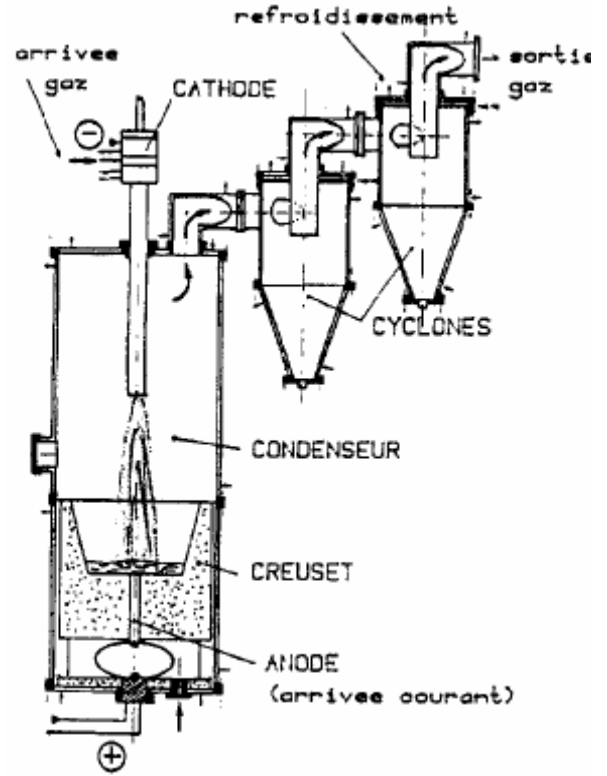

Figure 5. *Schéma du réacteur plasma de métallurgie extractive et métallurgie d'affinage 800 kW (1981)[11,12]*

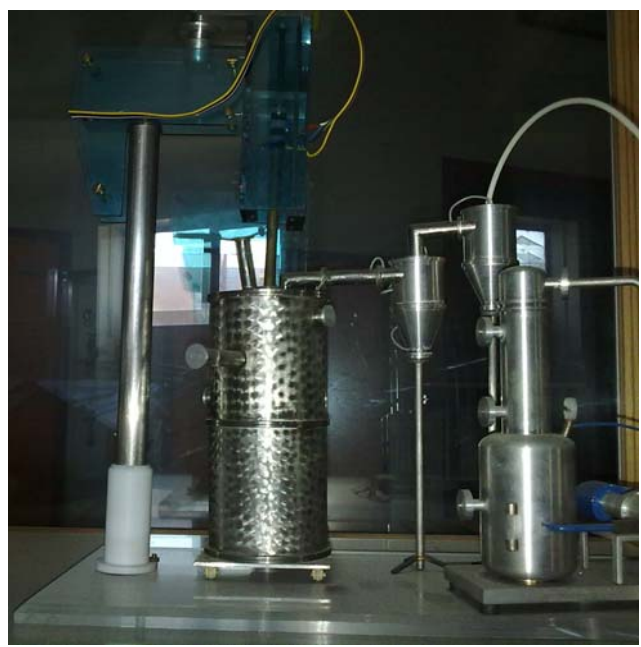

Figure 6- *Maquette du réacteur métallurgique à plasma d'arc transféré.*

### 3.3. Synthèse de poudres fines :

#### Génération d'aérosols (1981 À 1982)
Un générateur d'aérosols métalliques fut développé en collaboration avec les sociétés suédoises Marviken et SKF pour la vaporisation d'une anode constituée d'aciers et de métaux nobles. Le réacteur pilote était installé à Hoffors (Suède). Dans le cadre de cette collaboration internationale en matière de sûreté nucléaire (à laquelle participait le CEA), il s'agissait de simuler la fusion d'un cœur de réacteur nucléaire ("syndrome chinois") et d'observer la déposition dans les circuits, des aérosols résultant de cette fusion.

#### Synthèse du nitrure d'aluminium
La synthèse du nitrure d'aluminium, par vaporisation puis condensation réactive de l'aluminium en plasma d'azote a fait l'objet d'une thèse, d'un brevet ANVAR N° 86-04-108 et d'une concession d'option sur licence à COMAPEL filiale du groupe SFPO (Société du Ferromanganèse Paris-Outreau.). En effet le nitrure d'aluminium outre ses propriétés thermomécaniques, offre deux caractéristiques intéressantes : sa faible mouillabilité par les métaux liquides, qui permet de l'utiliser à la confection des creusets métallurgiques, et sa bonne conductivité thermique, qui est exploitée dans la réalisation de supports diélectriques qui évacuent bien la chaleur (supports de microcircuits électroniques).
La technologie mise au point fut adaptée à la production de poudres métalliques sphériques utilisées en métallisation.

### 3.4. Réacteur en lit fluide chauffé par plasma

**a) Synthèse de céramiques** (1985 À 1988)
- Utilisation de réacteurs à lit fluide chauffés par plasma (Voir figure 5) : fut la première réalisation de pilotes destinés à la carbonitruration d'alumine. La cible était la production de céramiques thermomécaniques transparentes aux ondes électromagnétiques centimétriques et la fabrication de radômes. Les mêmes réacteurs pouvaient être utilisés pour densifier des poudres céramiques obtenues par atomisation. (contrat BRITE, dossier de brevet EDF).

**b) Destruction de rejets industriels fluorés** (1985 À 1988)

Une technologie spécifique de destruction des déchets fluorés (pyralène) fut développée dans le cadre de l'ARC PIRSEM « Rejets industriels » du CNRS dirigée par Jacques AMOUROUX.

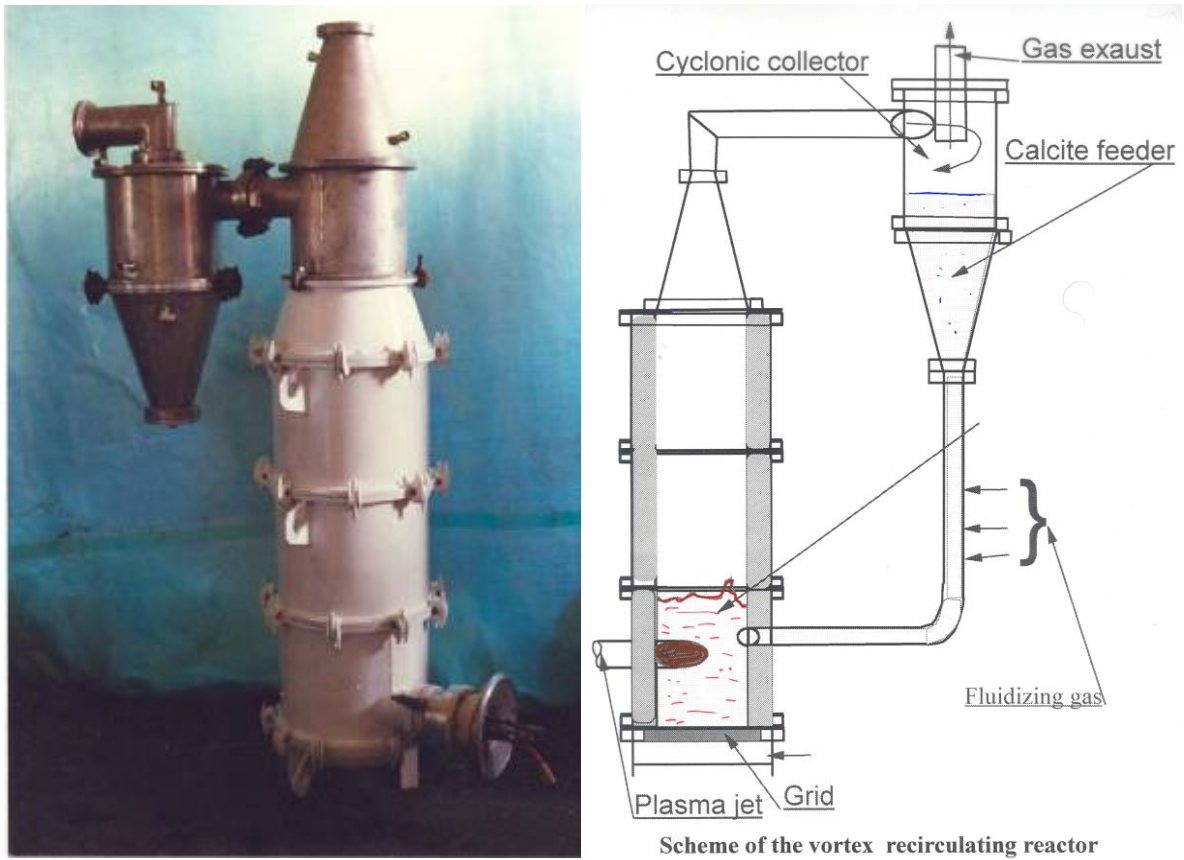

Figure 7. *Réacteur de synthèse ou de pyrolyse à lit fluide chauffé par plasma.*

### 3.5. Maquettage et étude paramétrique de la torche aérospatiale utilisée sur les hauts fourneaux (1988 à 1996)

Les hauts fourneaux métallurgiques sont des réacteurs en lit fixe qui exigent que leur charge reste poreuse, ce qui est assuré par l'utilisation de combustible sous la forme de coke, lequel présente une grande résistance mécanique à l'écrasement. Or le coke est devenu rare et n'est plus un sous-produit de la production de gaz de ville, enfin la pression écologique interdit de fait la construction de nouvelles cokeries. C'est pourquoi l'économie en coke peut être sensible en réchauffant par des torches plasmas les gaz au niveau des vents du haut fourneau.

C'est ainsi que sur le site de la SFPO à Boulogne mais aussi celui de Peugeot-Citroën à Sept-Fons des torches plasmas de type Aérospatiale furent expérimentées par EDF.

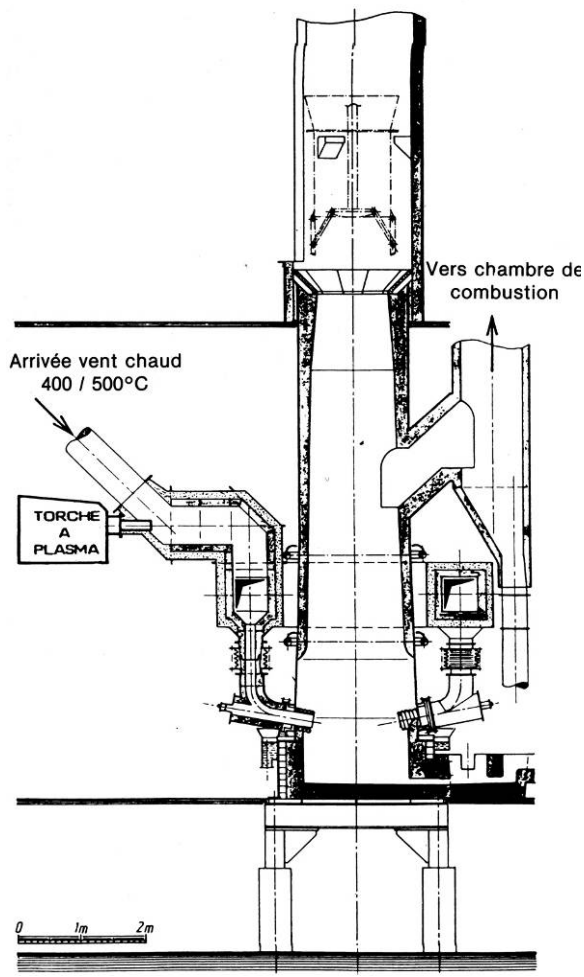
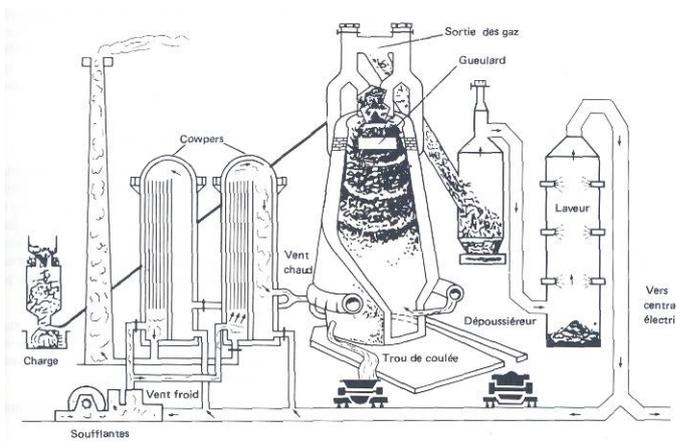
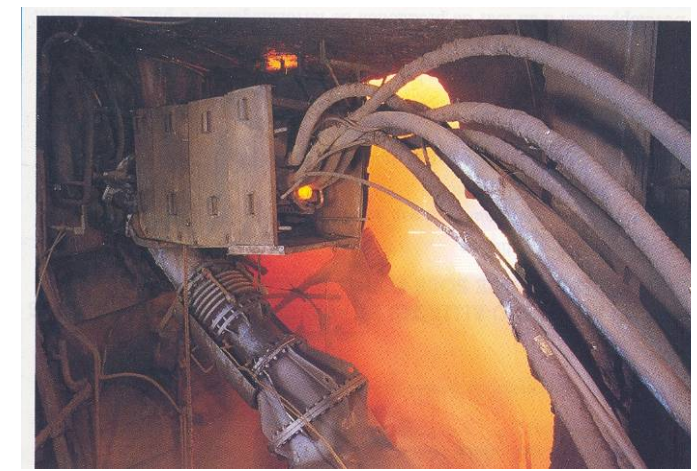

Figure 8 – *Cubillot ou bas fourneau de refusions de ferrailles à l'usine **Peugeot – Citroën** de Sept-fons (o3)[13]*

Figure 9 - *Schéma de principe des hauts fourneaux en métallurgie du ferromanganèse. Sur le site de Boulogne-Outreau (62) neuf torches de 2MW (comme celle ci-dessus) réchauffaient les vents chauds.*

Le laboratoire étudia pour EDF, le fonctionnement d'une torche à plasma d'arc (Plasma Energy Corporation) en chambre pressurisée dans ses configurations de plasma soufflé et plasma d'arc transféré (Voir Figure 7).

Une représentation paramétrique de fonctionnement fut établie ainsi que son domaine de validité par analyse dimensionnelle furent recherchées, en comparant les résultats obtenus avec deux torches *a priori* différentes, l'une de 2 MW réalisée par la Société Aérospatiale, l'autre de 0,2 MW construite par Plasma Energy Corporation (PEC). Ces études de caractéristiques électriques et thermiques de l'arc plasma, pour des arcs en plasma d'argon, d'azote ou d'air, ont permis de développer une série de corrélations utilisées actuellement par les industriels.

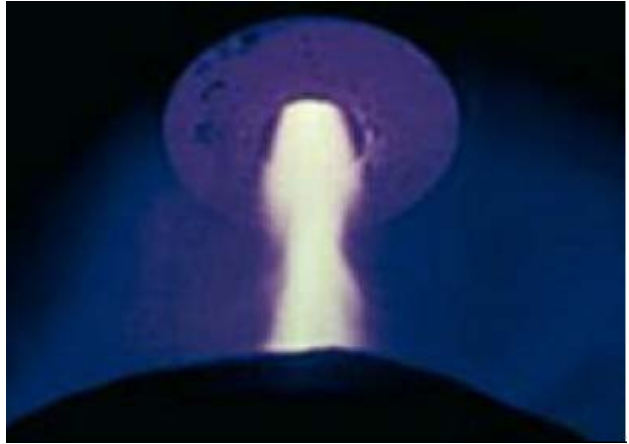

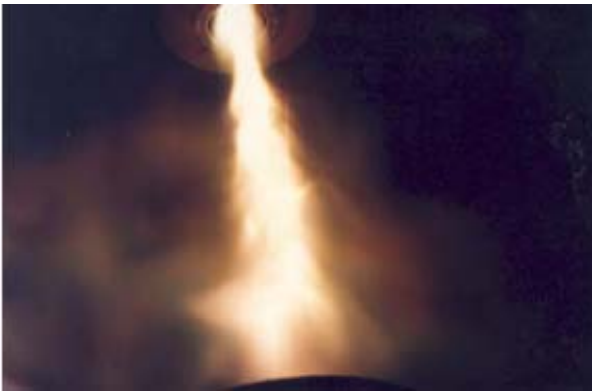
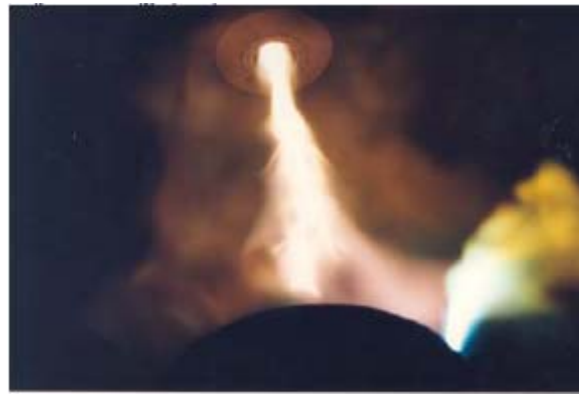

*Figure 10. Schémas des arcs transférés obtenu à partir de la torche PEC (Plasma Energy Corporation a subsidiary of the First Mississippi Corporation) en régime en régime anodique (gauche) et régime erratique (droite)*

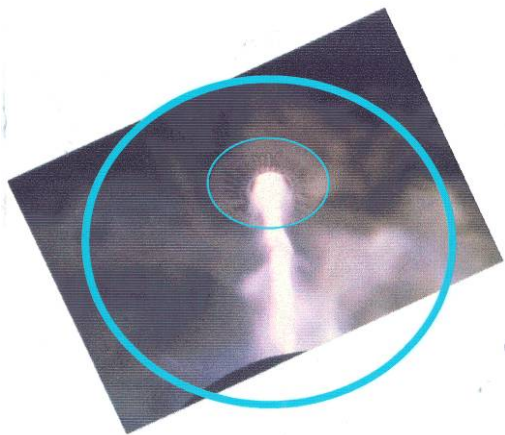
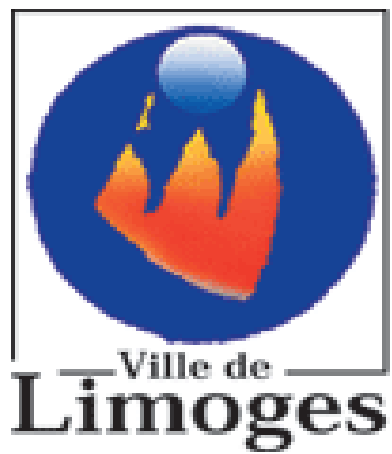

Figure 11 - *Schéma de principe du recadrage par l'artiste d'une des photographies de la figure 7.*

Figure 12 - *Logo officiel de la ville de Limoges*

## 4. Conclusions

En 2011, le paysage industriel et technologique français et mondial est totalement recomposé. La plupart des grosses entreprises industrielles présentes en 1977 ont disparu soit en nom soit en fait. Ainsi le groupe sidérurgique français **Usinor** fondé en 1948 a fusionné le 18 février 2002 avec l'espagnol **Aceralia** et le luxembourgeois **Arbed** pour former le groupe européen **Arcelor**. Au début de l'année 2006, le groupe néerlandais **Mittal** (d'origine indienne) lança une OPA hostile sur **Arcelor** qui aboutit à une fusion des deux groupes en juin 2006 pour former **Arcelor-Mittal**.

De même le groupe **Pechiney** (PUK) a été absorbé par **Alcan**, lequel a été racheté en 2007 par Rio Tinto pour devenir Rio Tinto Alcan. Menacé, en novembre 2007, d'une OPA hostile Rio Tinto Alcan. En août 2010, une nouvelle société, **Alcan EP**, est créée, regroupant l'activité produits usinés, c'est-à-dire les anciennes activités de Pechiney dans les produits en aluminium pour l'aéronautique, l'automobile ou l'industrie. Alcan EP compte 70 sites et 11 000 salariés dans le monde, dont 5000 en France. Alcan EP a vocation à revenir en Bourse[1].

**Electricité de France**, privée de son monopole de transport sur le territoire nationale se désintéresse de la recherche pour tenter une aventure, européenne voire mondiale, de conquête de marchés.

Chacun de ses regroupements ou réorientation se caractérise par l'abandon d'activités, voire de compression de personnel et pose donc le délicat problème de la mémoire industrielles et de la sauvegarde des « savoir-faire ».
Or en 2011 les plasmas thermiques industriels représentent un volume d'activités industrielles croissant sur trois principaux champs d'activité.
- La métallurgie (sidérurgie) avec la chauffe des cubilots de fonderies (Peugeot S.A. voir Figure 6a) et le maintien en température des poches et des sas de coulée continue.
- La destruction et recyclage de rejets ainsi l'amiante (EuroPlasma) et les organofluorés.
- Les dépôts par projection thermique

C'est pourquoi dans le cadre de ces journées nous sollicitons le témoignage des acteurs de cette aventure ainsi les collaborations sur des projets de même nature s'il y en a, sinon des témoignages, des documents, des participations actives au comité de pilotage du projet.

**Remerciements**

# I. Thèses soutenues sur le réacteur plasma.

**Mohammed TELLAT** *Contribution à l'étude théorique et expérimentale de la synthèse de nitrure et d'oxynitrure en réacteur à plasma d'arc à la pression atmosphérique.* Thèse soutenue le 28 Mars 1986, Limoges.

**Bouabdelli ABOULKASSIM** *Contribution à l'étude théorique et à la simulation d'un réacteur en lit fluidisé chauffé par plasma.* Thèse soutenue le 10 Juillet 1988, Limoges.

**Marie Françoise LERROL** *Étude et réalisation d'un réacteur plasma pilote de 1MW destiné à la métallurgie du manganèse.* Thèse soutenue le 23 Juin 1989, Limoges.

**Mohammed AFIFI** *Simulation tridimensionnelle transitoire de l'histoire thermique de matériaux chauffés par un jet plasma : méthode des éléments finis. Applications : traitement de surface.* Thèse soutenue le 8 Novembre 1991, Limoges.

**Jean François BRILHAC** *Contribution à l'étude statique et dynamique de torches plasma vortex stabilisées par vortex,* Thèse soutenue le 17 décembre 1993, Limoges. **_Primée en 1994 par le Club Enseignement Électrothermie._**

**David RIGOT** *Étude par la signature acoustique de l'érosion d'une tuyère de torche plasma*
Travaux codirigés avec les professeurs P. Fauchais et J. F. Coudert Thèse soutenue le 24 novembre 2003

**Julie CEDELLE** (soutenance décembre 2006) *Caractérisation multi échelles structurales et thermiques de dépôts de zircone pour SOFC et barrières thermiques*.

**Khalid. FATAOUI** (soutenance décembre 2007) *Caractérisation des propriétés thermomécaniques d'un dépôt en construction : étude de la propagation des fissures induites par une gouletette de matériau étranger.* Thèse en Cotutelle U. de Limoges et U. Chouaib Doukkali d'El Jadida (Maroc) codirigé par le Pr. Abdelmajid Belaphal

**Fadhel BENETTOUIL** (soutenance décembre 2007) *Modèle simplifié de la construction d'un dépôt par projection plasma et validation expérimentale*

# II. - LISTE DES PUBLICATIONS liées à l'activité Réacteurs Plasmas

## A. - PERIODIQUES

/A1/ **A. ADISSIN, M. BILLY, P. FAUCHAIS, P. GOURSAT, Ch. MARTIN, B. PATEYRON,** Etude par thermographie infrarouge du facteur d'émission de l'oxynitrure de silicium fritté : influence des ajouts et de la porosité. *C.R.A.S. Paris, 281B, (1975), 547.*

/A2/ **B. PATEYRON, A. ADISSIN, Ch. MARTIN, P. FAUCHAIS,** Etude par thermographie infrarouge du facteur d'émission de pièces céramiques en oxynitrure de silicium. *Revue Générale de Thermique, 172, (1976), 313*

/A3/ **A. ADISSIN, M. BILLY, P. FAUCHAIS, P. GOURSAT, Ch. MARTIN, B. PATEYRON,** Emissivity of silicon oxynitride ceramics by infrared thermography. *J. British Ceram. Soc., 32, (1976), 17.*

/A4/ **F. KASSABJI, B. PATEYRON, J. AUBRETON, M. BOULOS, P. FAUCHAIS.**
Conception d'un four à plasma de 0,7 MW pour la réduction des oxydes de fer. *Rev. Int. des Hautes Temp. et Réfract., 18, (1981).*

/A5/ **J. AUBRETON, B. PATEYRON, P. FAUCHAIS,** Les fours à Plasma. *Rev. Int. Hautes Temp. et Réfract., 18, (1981), 293.*

/A6/ **P. FAUCHAIS, A. et M. VARDELLE, J.F. COUDERT, B. PATEYRON,** State of the art in the field of plasma spraying and of extractive metallurgy with transferred arc: modelling, measurement, comparison between both, applications and developments, *Pure and Applied Chemistry, 57, (9), (1985), 1171.*

/A8/ **M.F. LERROL, B. PATEYRON, G. DELLUC, P. FAUCHAIS,** Etude dimensionnelle de l'arc électrique transféré utilisé en réacteur plasma, *Rev. Int. Hautes Temp. et Réfract., 24, 1988, 93-104.*

/A9/ **L. JESTIN, B. PATEYRON, P. FAUCHAIS,** Dimensionless study of experimental pressure related behavour for two powerful vortex blown arc and vortex-stabilzed power plasma torches, *High-Temp. Dust Laden Jets, pp. 211-231 (1989) Solonenko and Fedorchenko (Eds).*

/A10/ **B. PATEYRON, G. DELLUC, M.F. LERROL, P. FAUCHAIS,** Etude du fonctionnement d'un plasma d'arc électrique transféré utilisé en métallurgie extractive. *Rev. Int. Hautes Temp. et Réfract., 26, 1990, 1-7.*

/A11/ **J.F. BRILHAC, B. PATEYRON, P. FAUCHAIS, A. BOUVIER, P. PASQUINI, L. JESTIN**
Dimensionless relationships to calculate arc characteristics in a vortex D.C. plasma torch, *Journal of High Temperature Chemical Processes, Colloque, supplément au n°3, 1 (1992) pp 557-562*

/A12/ **J.F. BRILHAC, B. PATEYRON, J.F. COUDERT, P. FAUCHAIS, P. PASQUINI, A. BOUVIER, L. JESTIN,** Diagnostics of the dynamic behavior of the electric arc in a plasma torch *Journal of High Temperature Chemical Processes, Colloque, supplément au n°3, 1 (1992) pp 421-428*

/A13/ **M. AFIFI, B. PATEYRON, G. DELLUC, P. FAUCHAIS,** Modélisation tridimensionnelle de l'histoire thermique de matériaux chauffés par un jet plasma. Application à la projection plasma et aux traitements thermiques *Journal of High Temperature Chemical Processes, Colloque, supplément au n°3, 1 (1992) pp 397-402*

/A14/ **B. PATEYRON, G. DELLUC, M.F. ELCHINGER, P. FAUCHAIS,** Study of the behaviour of the heat conductivity and other transport properties of a simple reacting system: $H_2$-Ar and $H_2$-Ar-Air. Dilution effect in spraying process at atmospheric pressure *Journal of High Temperature Chemical Processes, Colloque, supplément au n°3, 1 (1992) pp 325-332*

/A15/ **B. PATEYRON, G. DELLUC, M.F. ELCHINGER, P. FAUCHAIS,** Thermodynamic and transport properties of Ar-$H_2$ and Ar-$H_2$-Air plasma gases used for spraying at atmospheric pressure *Plasma Chemistry Plasma Processing, Colloque, supplément au n°3, 1 (1992) pp 325-332*

/A16/ **J.F. BRILHAC, B. PATEYRON, P. FAUCHAIS,** Investigation of the thermal characteristics of d.c. vortex plasma torches *High Temp. Chem. Processes, 3 (1994) pp 419-425*

/A17/ **J.F. BRILHAC, B. PATEYRON, J.F. COUDERT, P. FAUCHAIS, A. BOUVIER,** Study of the dynamic and static behavior of DC vortex plasma torches : Part 1 : Button type cathode *Plasma chemistry and plasma processing, 15 (1) (1995) pp 257-277*

/A18/ **J.F. BRILHAC, B. PATEYRON, J.F. COUDERT, P. FAUCHAIS, A. BOUVIER,** study of the dynamic and static behavior of DC vortex plasma torches : Part 2 : Well-type cathode *Plasma chemistry and plasma processing, 15 (1) (1995) pp 231-255*

/A19/ **W.L.T. CHEN, J. OBERLEIN, E. PFENDER, B. PATEYRON, G. DELLUC, M.F. ELCHINGER, P. FAUCHAIS,** Thermodynamic and transport properties of argon/helium plasmas at atmospheric pressure *Plasma chemistry and plasma processing, 15 (3), sept. 1995, pp 559-579*

/A20/ **B. PATEYRON, M.F. ELCHINGER, G. DELLUC, P. FAUCHAIS,** Sound velocity in different reacting thermal plasma coatings *Plasma Chemistry Plasma Processing 16 (1) (1996) 39-57*

/A21/ **P. FAUCHAIS, J.F. COUDERT, B. PATEYRON,** La production de plasmas thermiques. *Revue Générale de Thermique, 35, 416, 1996, 543-560*

/A22/ **S. JANISSON, A. VARDELLE, J.F. COUDERT, E. MEILLOT, B. PATEYRON, P. FAUCHAIS,** Plasma spraying using Ar-He-H2 gas mixtures *J. Thermal Spray Technology, 8, 1999, 545-552*

/A23/ **G. MARIAUX, P. FAUCHAIS, A. VARDELLE, B. PATEYRON,** Modelling of the plasma spray process: from powder injection to coating formation *J. High Temp. Mat. Processes, 5, 2001, 61-85*

/A24/ **C. ALEMANY, C.TRASSY, B. PATEYRON, K.-I LI, Y. DELANNOY,** Refining of metallurgical-grade silicon by inductive plasma. *Solar Energy Materials and Solar Cells (2002), 72(1-4), 41-48.*

/A25/ **D. RIGOT, G. DELLUC, B. PATEYRON, J.F. COUDERT, P. FAUCHAIS, J. WIGREN**, Transient evolution and shifts of signals emitted by a d.c. plasma gun (type PTF4), *High Temp. Mat. Processes, 2, 2003, 175-185*

/A26/ **M. BOUNEDER, M. EL GANAOUI, B. PATEYRON, P. FAUCHAIS,** Thermal modelling of composite iron/alumina particles sprayed under plasma conditions. Part I : pure conduction, *J. High Temp. Mat. Processes, 7, 2003, 547-558*

/A27/ **G. LECOMTE, B. PATEYRON B AND P. BLANCHART,** Experimental study and simulation of a vertical section mullite-ternary eutectic (985 C) in the SiO2–Al2O3–K2O system, *Materials Research Bulletin, 3, 2004, 1469-1478*

/A28/ **S. DYSHLOVENKO, B. PATEYRON, L. PAWLOWSKI AND D. MURANO ,** Numerical simulation of hydroxyapatite powder behaviour in plasma jet, *Surface and Coatings Technology, 179, 1 , 2004, 110-117*

## B. - ACTES DE CONGRES

/B1/ **F. KASSABJI, J. AUBRETON, B. PATEYRON, P. FAUCHAIS,** Modélisation d'un four à plasma de forte puissance. *Congrès de la S.F.P., Toulouse, Juin, (1979).*

/B2/ **F. KASSABJI, J. AUBRETON, B. PATEYRON, P. FAUCHAIS, J. AMOUROUX, D. MORVAN,** Technical and economical studies for metal production by Plasma-Steel-Making application, *4th International Plasma Symposium on Plasma Chemistry, Zurich, August (1979), Proceedings p.349.*

/B3/ **P. FAUCHAIS, M. VARDELLE, B. PATEYRON,** La métallurgie extractive dans les fours à plasma et les traitements de surface par projection plasma. *9ème Congrès International UIE9, Cannes, 20.24 Octobre (1980), Proceeding.*

/B4/ **B. PATEYRON, J. AUBRETON, F. KASSABJI, P. FAUCHAIS,** Some new design of reduction plasma furnaces including the hollow cathode system, electrical transfer to the bath and falling film. *5th International Symposium on Plasma Chemistry University of Edinburg. (ed. Waldie) (1981).*

/B41/ **S. BERNARD, P. FAUCHAIS, J. JARRIGE, J.P. LECOMPTE, B. PATEYRON,** Lead and zinc evaporation competition in the treatment of fly ashes model by plasma transferred arc. *Proceedings 15th International Symposium on Plasma Chemistry, Orleans, 9-13 juillet 2001, A. Bouchoule, J.M. Pouvesle, A.L. Thomann, J.M. Bauchire, E. Robert Eds., Vol. V, 2029-2036*

/B42/ **C.ALEMANY, K.I. LI, Y. DELANNOY, B. PATEYRON, P. PROULX, D. MORVAN, C. TRASSY,** Plasma refining of metallurgical silicon : thermodynamic and chemical aspects, *Proceedings of the 7th European Conference on Thermal Plasma Processes, Strasbourg, 18-21 Juin 2002, Progress in Plasma Processing of Materials*, P. Fauchais Ed., Begell House, N.Y. (USA), 2003, 717-722

/B43/ **B. PATEYRON, P. PROULX, C. TRASSY,** Effect of electric fields on the non-equilibrium in an inductively coupled plasma *Progress in Plasma Processing of Materials 2003, E MRS-IUMRS-ICEM 2002, TPP7 Thermal Plasma Processes, Strasbourg, 18-21 juin 2002, P. Fauchais Eds., Begell House, N.Y.(USA), 2003, 717-722*

### Autres

/C1/ **D. YEROUCHALMI, S. DALLAIRE, B. PATEYRON**, Production de poudre de silice pure ultrafine par fusion suivie de vaporisation dans un four tournant chauffé par plasma d'arc. 3ème Colloque International sur les Fenêtres Électromagnétiques. Paris, Sept. 1975 (Ed. D.M.A.).

/C2/ **B. PATEYRON, G. DELLUC, M.F. LERROL, P. FAUCHAIS,** Étude de l'arc électrique transféré utilisé en réacteur plasma, Journées Ampères (1986) Proc. p 55 (ed) SEE 48 rue de la Procession, Paris.

/C3/ **M.F. ELCHINGER, B. PATEYRON, P. FAUCHAIS, A. VARDELLE,** Calculation of thermodynamic and transport properties of Ar-H2-Air plasma, comparison with simple mixing rules.
13th International Symposium on Plasma Chemistry, Symposium Proceedings, Supplement, August 18-22, Beijing (Chine), 1997-2003

/C4/ **B. PATEYRON, M.F. ELCHINGER, G. DELLUC, P. FAUCHAIS**, Thermodynamic and transport properties of the plasma ternary mixture Ar-He-N2, 13th International Symposium on Plasma Chemistry, Symposium Proceedings, Supplement, August 18-22, Beijing (Chine), 2004-2009

/C5/ **B. PATEYRON, G. DELLUC, M. EL GANAOUI, A. VARDELLE, M. VARDELLE, J. F. COUDERT AND P. FAUCHAIS.** Transfert de chaleur et de masse dans les jets plasmagènes impactants. Journée thématique CEA/SFT : transferts de chaleur et de masse dans les jets. Paris, 14 Mars 2001.

/C6/ **P. FAUCHAIS, A. VARDELLE, A. DENOIRJEAN, B. PATEYRON, M. EL GANAOUI**, Alumina splats formation layering: thermal history of coating formation resulting residual stresses and coating microstructure The International Thermal Spray Conference – Advancing Thermal Spray in the 21st Century, Singapour, 28-30 Mai 2001, New Surfaces for a New Millenium, C. Berndt, A. Khor, E. Lugscheider Eds., ASM Pub., Ohio (USA), 865-874

/C7/ **G.DELLUC, G. MARIAUX, A. VARDELLE, P. FAUCHAIS, B. PATEYRON**, A numerical tool for plasma spraying. Part I : modelling of plasma jet and particle behavior 16th International Symposium on Plasma Chemistry, ISPC16, Taormina (Italie), 22-27 Juin 2003, R. D'Agostino, P. Favia, F. Fracassi Eds., Univ. of Bari, CD-R, 6 pages

/C8/ **G.DELLUC, L. PERRIN, H. AGEORGES, P. FAUCHAIS, B. PATEYRON,** A numerical tool for plasma spraying. Part II : model of statistic distribution of alumina multi particle powder, 16th International Symposium on Plasma Chemistry, ISPC16, Taormina (Italie), 22-27 Juin 2003, R. D'Agostino, P. Favia, F. Fracassi Eds., Univ. of Bari, CD-R, 6 pages

/C9/ **M.EL GANAOUI, M. BOUNEDER, B. PATEYRON, P. FAUCHAIS**, Thermal modelling of multilayer sprayed particles under plasma conditions, 16th International Symposium on Plasma Chemistry, ISPC16, Taormina (Italie), 22-27 Juin 2003, R. D'Agostino, P. Favia, F. Fracassi Eds., Univ. of Bari, CD-R, 6 pages

/C10/ **J.CEDELLE, S. BANSARD, C. ESCURE, M. VARDELLE, B. PATEYRON, P. FAUCHAIS**, Experimental investigation of the splashing processes at impact in plasma sprayed coating formation, 16th International Symposium on Plasma Chemistry, ISPC16, Taormina (Italie), 22-27 Juin 2003, R. D'Agostino, P. Favia, F. Fracassi Eds., Univ. of Bari, CD-R, 5 pages

### C. AFFICHES, montage video, etc...

/D1/ **B. PATEYRON,** Réacteurs plasma de métallurgie extractive dans l'équipe thermodynamique et plasma de l'UA 320 CNRS/Université de Limoges. *Montage Vidéo 12' (français) présenté au Congrès de l'Union Internationale d'Electricité, Stockholm, SUEDE, 1984.*

/D2/ **B. PATEYRON, P. FAUCHAIS,** Metallic aerosols production in arc plasma reactor. *Gordon Conf. on Plasma chemistry, Tilton (U.S.A.), aug. 1984.*

/D3/ **B. PATEYRON, P. FAUCHAIS,** Aluminum nitride synthesis in plasma reactor. *Gordon Conf. on Plasma chemistry, Tilton (U.S.A.), aug. 1986.*

microstructure *The International Thermal Spray Conference – Advancing Thermal Spray in the 21st Century, Singapour, 28-30 Mai 2001*

/D49/ **S. BERNARD, P. FAUCHAIS, J. JARRIGE, J.P. LECOMPTE, B. PATEYRON**
Lead and zinc evaporation competition in the treatment of fly ashes model by plasma transferred arc *15th International Symposium on Plasma Chemistry, ISPC15, Orléans, 9-13 Juillet 2001*

/D50/ **B. PATEYRON, G. DELLUC,** T&Twinner base de calculs des propriétés de transport des gaz plasmas *Journées CEA-Centre Européen de la Céramique, Limoges, 13 Décembre 2001*

/D51/ **B. PATEYRON, A. VARDELLE, M. VARDELLE, G. DELLUC, M. EL GANAOUI, P. FAUCHAIS, D. GOBIN,** Etude numérique et expérimentale de la construction d'un dépôt par projection plasma : de la lamelle au dépôt, *Journées CEA-Centre Européen de la Céramique, Limoges, 13 Décembre 2001*

/D52/ **G. DELLUC, B. PATEYRON, H. AGEORGES, M. EL GANAOUI ET P. FAUCHAIS,** Recherche d'une méthode numérique rapide d'évaluation de l'histoire thermique et des changements de phase dans une sphère mono-matériau, *Congrès Français de Thermique SFT 2001, Nantes 29-31 mai 2001. (poster)*

/D53/ **B. PATEYRON, G. DELLUC,** T&TWinner base de calcul des propriétés thermochimiques et de transport des gaz plasmas, *Journée Scientifique SFC, section Centre-Ouest "Matériaux à propriétés spécifiques : de la conception aux applications", Limoges, 18 Janvier 2002*

/D54/ **C. ALEMANY, C. TRASSY, B. PATEYRON, K-I LI, Y. DELANNOY,** Plasma refining of metallurgical silicon: thermodynamic and chemical aspects *E MRS-IUMRS-ICEM 2002, TPP7 Thermal Plasma Processes, Strasbourg, , 18-21 juin 2002*

/D55/ **B. PATEYRON, P. PROULX, C. TRASSY,** Effect of electric fields on the non-equilibrium in an inductively coupled plasma *E MRS-IUMRS-ICEM 2002, TPP7 Thermal Plasma Processes, Strasbourg, 18-21 juin 2002*

/D56/ **D. RIGOT, B. PATEYRON, J.F. COUDERT, P. FAUCHAIS, J. WIGREN,** Evolutions et dérives des signaux émis par une torche à plasma à courant continu (type PTF4) – *6ème Journées d'études sur les fluctuations d'arc – LAEPT, Université Blaise Pascal – Clermont Ferrand, 17 et 18 Mars 2003.*

/D57/ **J. CEDELLE, S. BANSARD, C. ESCURE, M. VARDELLE , B. PATEYRON, P. FAUCHAIS** Experimental investigation of the splashing processes at impact in plasma sprayed coating formation *16$^{th}$ ISPC Taormina, Italy, June 22-27, 2003*

/D58/ **G. DELLUC, G. MARIAUX, A VARDELLE., P.FAUCHAIS, B. PATEYRON,** A numerical tool for plasma spraying, Part I: modelling of plasma jet and particle behaviour *16$^{th}$ ISPC Taormina, Italy, June 22-27, 2003*

/D59/ **G. DELLUC, L. PERRIN, H. AGEORGES, P. FAUCHAIS, B. PATEYRON,** A numerical tool for plasma spraying, Part II: Model of statistic distribution of alumina multi particle powder *16$^{th}$ ISPC Taormina, Italy, June 22-27, 2003*

/D60/ **M. EL GANAOUI, M. BOUNEDER, B. PATEYRON, P. FAUCHAIS,** Thermal modelling of multilayer sprayed particles under plasma conditions *16$^{th}$ ISPC Taormina, Italy, June 22-27, 2003*

/D61/ / **G. DELLUC, B. PATEYRON,** Echanges d'énergie et quantités de mouvement entre plasmas et particules, *Ateliers thématiques du réseau plasmas froids : nucléation, croissance et transport de nano-particules dans un plasma, 7 et 8 janvier 2003, Gif-sur-Yvette (Centre de formation du CNRS)*

/D62/ **C. KATSONIS, B. PATEYRON,** Calculs des propriétés thermodynamiques et de transport en plasmas froids, *Ateliers thématiques du réseau plasmas froids : Modélisation plasma : ECOMOD Le Tolosan, Boussens (31), 2-3 Juin 2003.*

/D63/ **DELLUC G., ELGANAOUI M., PATEYRON B., FAUCHAIS P. ,** Comparaison d'une méthode intégrale avec une méthode de différences finies dans l'évaluation de l'histoire thermique et des changements de phase dans une sphère mono matériau. *PIRMII Marrakech 2003*

/D64/ **DELLUC G., PATEYRON B.** Modélisation de jets à haute température et de leurs interactions avec les poudres, PRIPT, *Premières Rencontres Internationales sur la Projection Thermique Villeneuve d'Ascq (Lille), 4 et 5 décembre 2003*

/D65/ **M. BOUNEDER, M. EL GANAOUI, B PATEYRON, P. FAUCHAIS,** Computational heat transfer and phase change in composite (metal/ceramic) particle immersed in a plasma pool, *Proceedings of the International Thermal Spray Conference, ITSC2004, Osaka (Japon), 10-12 Mai 2004, Thermal Spray Solutions : Advances in Technology and Application, ASM International, CD-R, 5 pages*

/D66/ **G. DELLUC, L. PERRIN, H. AGEORGES, P. FAUCHAIS, B. PATEYRON**
Modelling of plasma jet and particle behavior in spraying conditions, *Proceedings of the International Thermal Spray Conference, ITSC2004, Osaka (Japon), 10-12 Mai 2004, Thermal Spray Solutions : Advances in Technology and Application, ASM International, CD-R, 6 pages*

/D66/ **J. CEDELLE, M. VARDELLE, B. PATEYRON, P. FAUCHAIS,** Experimental investigations of the splashing processes at impact in plasma sprayed coating formation, *Proceedings of the International Thermal Spray Conference, ITSC2004, Osaka (Japon), 10-12 Mai 2004, Thermal Spray Solutions : Advances in Technology and Application, ASM International, CD-R, 6 pages*

/D67/ **D. RIGOT, B. PATEYRON, J.F. COUDERT, P. FAUCHAIS, J. WIGREN,** Study of the erosion of electrodes in D.C; plasma torches by on line monitoring of the voltage and the sound, *Proceedings of the International Thermal Spray Conference, ITSC2004, Osaka (Japon), 10-12 Mai 2004, Thermal Spray Solutions : Advances in Technology and Application, ASM International, CD-R, 6 pages*

/D68/ **D. RIGOT, J.F. COUDERT, B. PATEYRON, P. FAUCHAIS, J. WIGREN,** Contribution à l'étude d'un modèle thermique lié à la stagnation du pied d'arc sur l'anode, dans un torche de projection plasma, *Actes du Congrès Français de Thermique – Transferts en Milieux Hétérogènes, Presqu'île de Giens, 25-28 Mai 2004, D. Gobin, M. Pons, G. Lauriat, P. Le Quéré Eds., Tome 1, 2004, 393-398*

/D69/ **J. CEDELLE, M. VARDELLE, B. PATEYRON, P. FAUCHAIS,** Etude expérimentale du phénomène de spashing à l'impact lors de la réalisation d'un dépôt par projection plasma, *Actes du Congrès Français de Thermique – Transferts en Milieux Hétérogènes, Presqu'île de Giens, 25-28 Mai 2004, D. Gobin, M. Pons, G. Lauriat, P. Le Quéré Eds., Tome 1, 2004, 405-410*

/D70/ **B. PATEYRON, K. FATAOUI, A.S. SYED, A. DENOIRJEAN, P. FAUCHAIS,** Comparaison de modèles de turbulence dans le modèle d'écoulement d'un jet plasma utilisé pour la projection de poudre métallique, *Actes du Congrès Français de Thermique – Transferts en Milieux Hétérogènes, Presqu'île de Giens, 25-28 Mai 2004, D. Gobin, M. Pons, G. Lauriat, P. Le Quéré Eds., Tome 1, 2004, 453-458*

/D71/ **M. BOUNEDER, M. EL GANAOUI, B. PATEYRON, P. FAUCHAIS,** Transfert de chaleur avec transition de phase dans une sphère composite (métal-céramique) immergée dans un gaz plasmagène, *Actes du Congrès Français de Thermique – Transferts en Milieux Hétérogènes, Presqu'île de Giens, 25-28 Mai 2004, D. Gobin, M. Pons, G. Lauriat, P. Le Quéré Eds., Tome 2, 2004, 575-580*

/D72/ **A. VARDELLE, G. MARIAUX, B. PATEYRON, M. EL GANAOUI, P. FAUCHAIS**
Modelling of the plasma spray process: from powder injection to coating formation *The European Material Conference TPP6 Thermal plasma processes Strasbourg 30 mai-2 juin 2000*

### D. CONTRIBUTION à des ouvrages.

Tables et calculs dans : **M. BOULOS, P. FAUCHAIS, E. PFENDER** Thermal Plasmas Fundamentals and Applications, *Vol 1 (Pub.) Plenum Press, NY and London, (1994), 452 pages*

**FAUCHAIS P., A. VARDELLE, M. VARDELLE, A. DENOIRJEAN, B. PATEYRON AND M. EL GANAOUI,** Formation and layering of alumina splats : thermal history of coating formation, resulting residual stresses and coating microstructure. *Thermal Spray 2001: New Surfaces for a New Millennium, (Ed.) C. C. Berndt, K. A. Khor, and E. F. Lugscheider, ASM International, Material Park, Ohio, USA, pp. 865-873, 2001.*

**PATEYRON B., VARDELLE A., EL GANAOUI M., G. DELLUC, P. FAUCHAIS,** 1D Modeling of coating formation under plasma spraying conditions: Splat cooling and laying. *Progress in plasma processing of materials, Ed. P. Fauchais, Begell House, p.519-526, 2001.*

### E. BREVETS et Dossiers de Valorisation ANVAR

**B. PATEYRON, M.F. ELCHINGER, G. DELLUC, J. AUBRETON,** Logiciel de calculs d'équilibres chimiques complexes : *TEM. Dossier de valorisation ANVAR n° 52. 356, (1986).*

**B. PATEYRON, E. LEJAMTEL, P. FAUCHAIS,** Lits fluides chauffés par plasma. *Dossier de brevet aux bons soins de la DER d'EDF (1988).*

**B. PATEYRON, E. LEJAMTEL, P. FAUCHAIS,** Electrode Plasma industrielle. *Dossier de brevet aux bons soins de la DER d'EDF (1988).*

**B. PATEYRON, E. LEJAMTEL, P. FAUCHAIS,** *Synthèse de poudres fines métalliques ou non en réacteur plasma.* **ANVAR -86-04-108-.** Contrat de Licence concédé à **COMAPEL n° 90 6196 00** du 30 Juillet 1991. (Durée 5 ans)

**J.F. BRILHAC, L.JESTIN, A. BOUVIER, G.TREILLARD, P. FAUCHAIS, B. PATEYRON,**
Torche à plasma modulaire. *EDF-Lavoix B 1636, 1994, France*

Bernard PATEYRON revendique en outre une contribution décisive au brevet suivant :

**M.D. NICOUD, J.M. LEGER, P.FAUCHAIS, A. GRIMAUD,** Mélanges ternaires pour projection plasma. *N° 90-04-307 du 04-04-90. Air Liquide*, F, CEE, USA, CN, Japon;

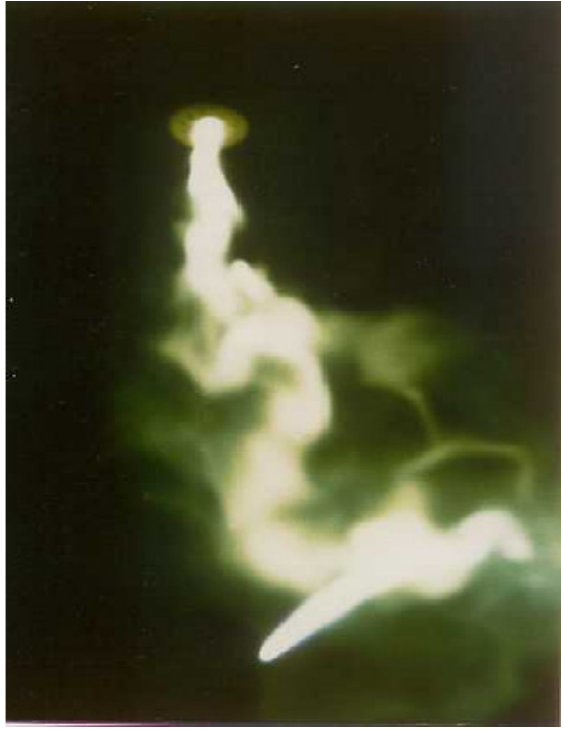

*Figure 13- Torche Plasma Energy Corporation sous flux d'air. (Arc long de un mètre, intensité de 300 ampères sous 800 volts). L'arc est erratique en zone anodique sur un creuset de graphite.*

**Références**